\documentclass[journal=jacs,manuscript=communication]{achemso}

\usepackage{chemformula} 
\usepackage[T1]{fontenc} 
\usepackage{color}
\usepackage[scr=rsfso]{mathalpha}
\usepackage{bm}
\usepackage{physics}
\usepackage[T1]{fontenc}
\usepackage[normalem]{ulem}
\usepackage{soul}

\author{Francesco Verdelli}
\email{f.verdelli@tue.nl}
\affiliation{Dutch Institute for Fundamental Energy Research, 5600HH Eindhoven, The Netherlands.}
\author{Yu-Chen Wei}
\email{y.c.wei@tue.nl}
\affiliation{Department of Applied Physics and Science Education, Eindhoven University of Technology, 5600MB Eindhoven, The Netherlands}
\author{Kripa Joseph}
\affiliation{Department of Chemical Engineering and Chemistry, Eindhoven University of Technology, 5600MB Eindhoven, The Netherlands}
\author{Mohamed S. Abdelkhalik}
\affiliation{Department of Applied Physics and Science Education, Eindhoven University of Technology, 5600MB Eindhoven, The Netherlands}
\author{Goudarzi Masoumeh}
\affiliation{Department of Applied Physics and Science Education, Eindhoven University of Technology, 5600MB Eindhoven, The Netherlands}
\author{Sven H.C. Askes}
\affiliation{Department of Physics and Astronomy, Vrije Universiteit Amsterdam, 1081HV Amsterdam, The Netherlands}%
\author{Andrea Baldi}
\affiliation{Department of Physics and Astronomy, Vrije Universiteit Amsterdam, 1081HV Amsterdam, The Netherlands}%
\author{E. W. Meijer}
\affiliation{Department of Chemical Engineering and Chemistry, Eindhoven University of Technology, 5600MB Eindhoven, The Netherlands}
\author{Jaime G\'omez Rivas}
\email{j.gomez.rivas@tue.nl}
\affiliation{Department of Applied Physics and Science Education, Eindhoven University of Technology, 5600MB Eindhoven, The Netherlands}

\title[Polaritonic Chemistry Enabled by Non-Local Metasurfaces]
  {Polaritonic Chemistry Enabled by Non-Local Metasurfaces}

\begin{document}


\begin{tocentry}

    \includegraphics[scale=0.5]{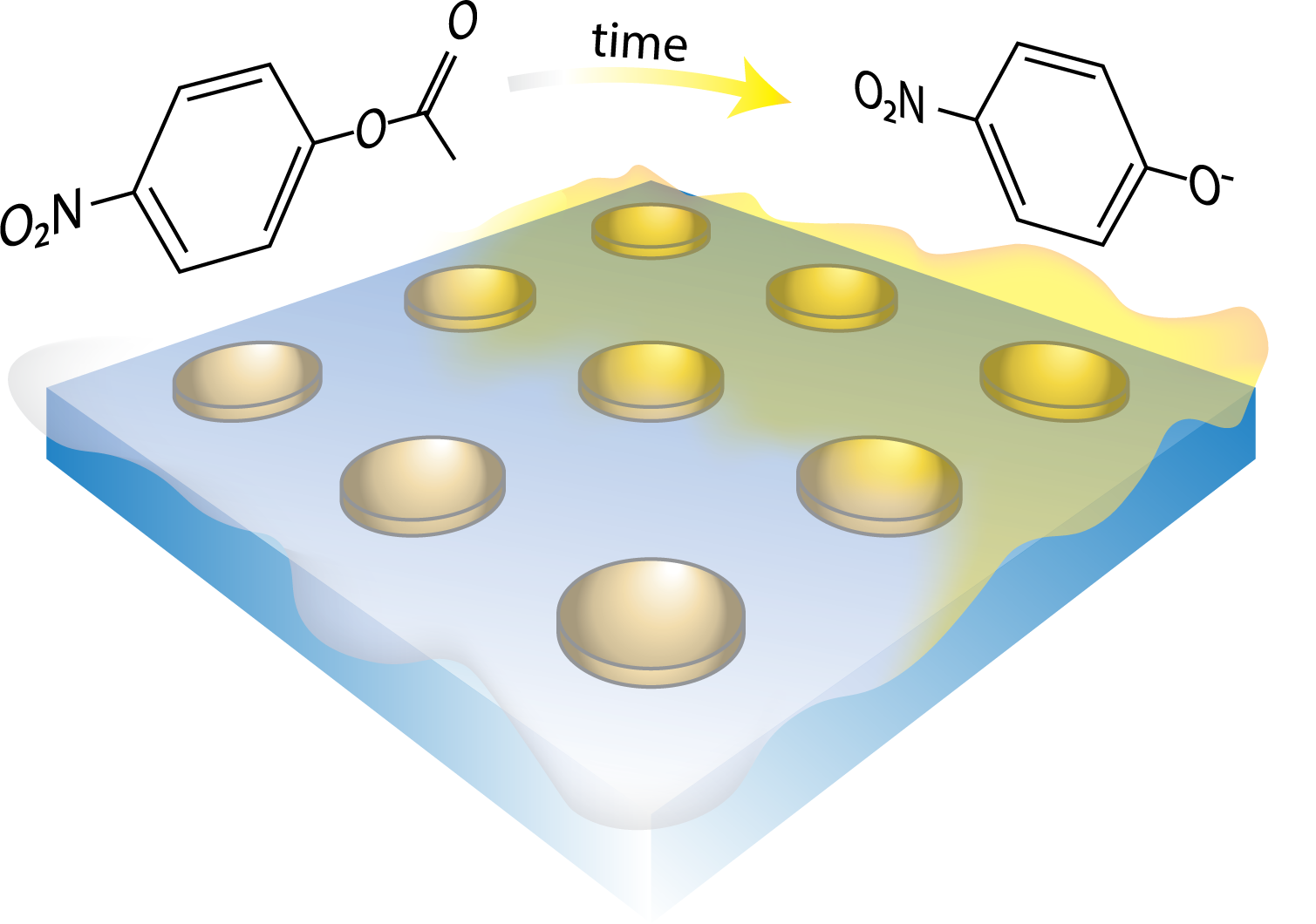}

\end{tocentry}


\begin{abstract}
Vibrational strong coupling can modify chemical reaction pathways in unconventional ways. Thus far, Fabry-Perot cavities formed by pairs of facing mirrors have been mostly utilized to achieve vibrational strong coupling. In this study, we demonstrate the application of plasmonic microparticle arrays defining non-local metasurfaces that can sustain surface lattice resonances as a novel tool to enable chemical reactions under vibrational strong coupling. We show that the solvolysis kinetics of \textit{para}-nitrophenyl acetate can be accelerated by a factor of 2.7 by strong coupling to the carbonyl bond of the solvent and the solute with a surface lattice resonance. Our work introduces a new platform to investigate and control polaritonic chemical reactions. In contrast to Fabry-Perot cavities, metasurfaces define open optical cavities with single surfaces, which removes alignment hurdles, facilitating polaritonic chemistry across large areas.
\end{abstract}

Catalyzing chemical reactions has a profound impact on society.\cite{gelle2019applications,norskov2008nature,swearer2016heterometallic,seh2017combining,wismann2019electrified,hartman2020operando,jacobson2020shining,wilson2020light,devasia2021control,candish2021photocatalysis}  In the pursuit of innovative catalytic approaches, vibrational strong coupling (VSC) has recently emerged as a powerful and non-intrusive tool for modifying chemical reactions. \cite{houdre2005early,torma2014strong,ebbesen2016hybrid,sidler2020polaritonic,nagarajan2021chemistry,li2022molecular} This strong interaction arises when the exchange of energy between photons and molecular vibrations exceeds the dissipation rates, leading to the emergence of two novel hybrid states, called lower and upper polaritons, in addition to vibrational dark states. Previous studies demonstrate that VSC can influence chemical processes,~\cite{hirai2020recent,nagarajan2021chemistry} including acceleration \cite{lather2019cavity,lather2020improving,george2021cavity} and suppression\cite{thomas2016ground,vergauwe2019modification,ahn2023modification} of reaction rates, as well as modifying product selectivity\cite{thomas2019tilting,pang2020role,sau2021modifying}. Various theoretical models have been developed to elucidate the fundamental mechanisms of VSC in chemical reactions \cite{feist2018polaritonic,yuen2019polariton,sanchez2022theoretical,fregoni2022theoretical,du2022catalysis,du2023vibropolaritonic,lindoy2023quantum}, opening avenues to explore their applicability and universality. All these reactions investigated under VSC belong to a burgeoning branch of chemistry known as polaritonic chemistry.\cite{fregoni2023polaritonic} 

So far, the preferred optical cavity for polaritonic chemistry has been the Fabry-Perot cavity \cite{simpkins2023control}. This cavity, owing to its straightforward fabrication and versatile tuning capabilities, offers a favorable enclosed system for optical mode confinement. However, slight misalignments between the two facing mirrors could introduce measurement errors and hinder reproducibility \cite{imperatore2021reproducibility,simpkins2021mode,wiesehan2021negligible}. 
To overcome these challenges, our focus shifts to non-local metasurfaces formed by arrays of plasmonic particles as an alternative means of light confinement on the surface, providing substantial field enhancements. These metasurfaces support highly tunable optical modes, called surface lattice resonances (SLRs), arising from the enhanced radiative coupling of the localized plasmon resonances of the individual scatterers mediated by the diffraction orders of the array.\cite{rodriguez2011coupling,wang2018rich,kravets2018plasmonic} It has already been demonstrated that SLRs can exhibit VSC with molecular vibrations in thin films.\cite{cohn2021infrared,brawley2021angle,verdelli2022chasing} Compared to Fabry-Perot systems, SLRs-based cavities show larger field enhancements and can be conveniently fabricated over extensive areas. Since the resonance of the metasurface is defined by the scatterers at a single interface, it is not necessary to align the two surfaces that define the Fabry-Perot cavity after fabrication. The combination of these characteristics makes metasurfaces promising contenders as a groundbreaking platform in polaritonic chemistry.

In this manuscript, we design a gold microparticle array that couples the C$=$O stretching mode of ethyl acetate (EtOAc) and \textit{para}-nitrophenyl acetate (PNPA) with SLRs.  Previous work by Lather et al.\cite{lather2019cavity} demonstrated accelerated kinetics in this solvolysis reaction through cooperative coupling of the carbonyl peak with the optical mode of a Fabry-Perot cavity. Our study confirms that the strong coupling between the C$=$O stretching mode and the SLR of the metasurface (Rabi splitting of 77 cm$^{-1}$) increases the reaction rate by a factor of 2.7 compared to the non-cavity reaction. Our findings introduce non-local metasurfaces as alternative platforms for catalyzing chemical reactions via VSC, offering new possibilities for industrial-level device configurations and catalytic processes.  
\begin{figure}[h]
    \centering
    \includegraphics[scale=0.6]{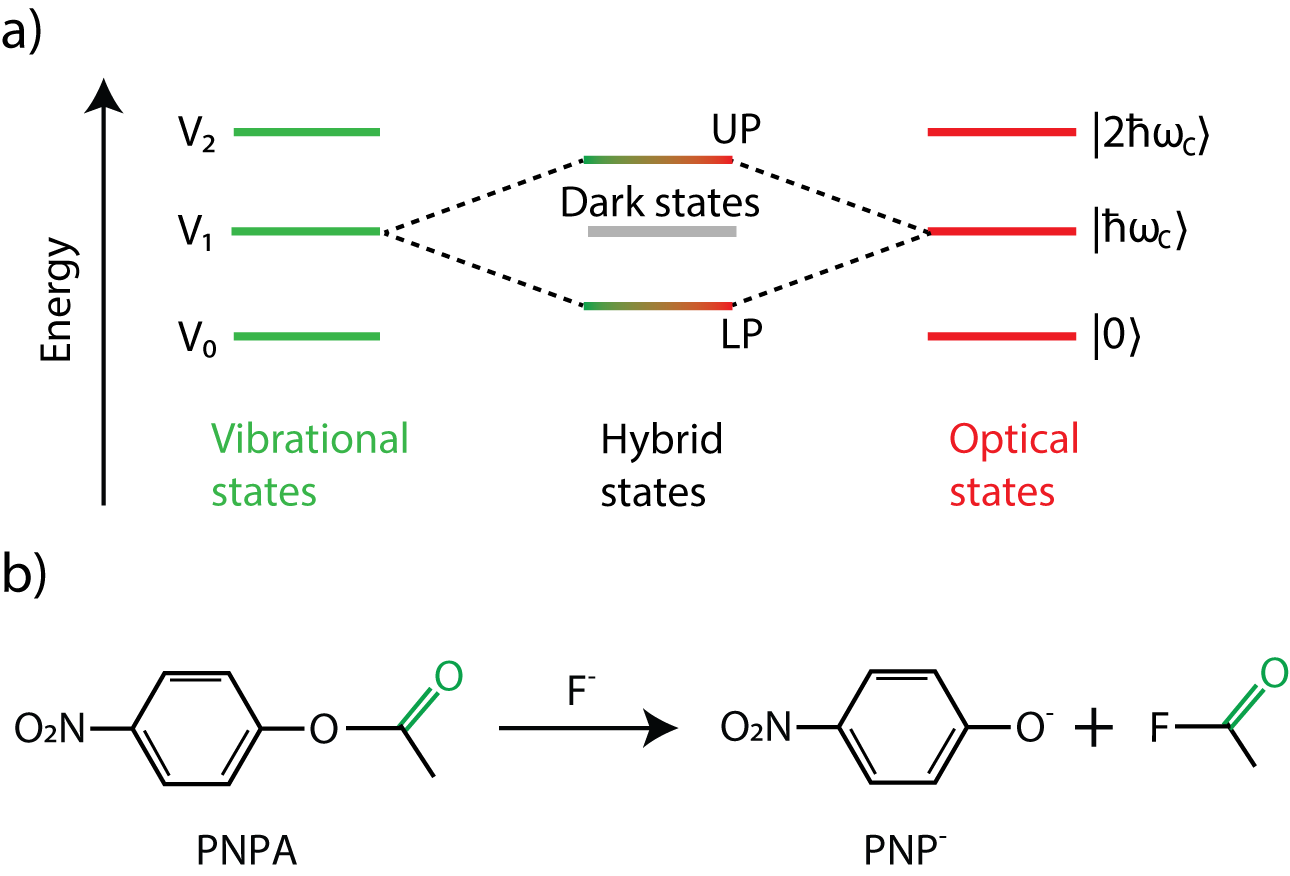}
    \caption{Schematic illustration of vibrational strong coupling and the reaction under investigation. a) Energy level diagram of the coupled system. The vibrational states are shown as green lines. The optical modes of the cavity are shown as red lines. The hybrid states are shown as green-red lines, and the vibrational dark states as a gray line.  b) Reaction equation of the solvolysis of \textit{para}-nitrophenyl acetate into the product \textit{para}-nitrophenoxide. The carbonyl peak involved in the reaction is highlighted in green.}
    \label{schematic}
    \label{fig:fig1}
\end{figure}

Figure~\ref{schematic}a depicts a schematic of energy levels during VSC between the optical modes (red) and the fundamental vibrational frequency of the C$=$O bond (green). The resulting hybrid states, upper polariton (UP) and lower polariton (LP), exhibit an energy separation known as Rabi splitting. The magnitude of the Rabi splitting is determined by three factors: the oscillator strength of the molecular vibration, the volume of the optical mode of the cavity, $V_{m}$, and the number of molecules $N$ within it. The dependence on the latter two factors is given by $\sqrt{\frac{N}{V_{m}}}$. Consequently, increasing the number of molecules within the optical cavity field and enhancing the cavity field confinement yields the maximum Rabi splitting. 

To study the effects of VSC on chemical reactions, we focus on the ester solvolysis reaction of PNPA (Figure~\ref{schematic}b). The reaction mechanism involves the nucleophilic attack of the carbonyl bond by F$^{-}$ ions, leading to the formation of \textit{para}-nitrophenolate (PNP$^{-}$) and acetyl fluoride (see SI section S1).\cite{park1961hydrolysis} The time evolution of the concentration of PNP$^{-}$ can be monitored in the visible range by recording the intensity of its absorbance peak \cite{lather2019cavity}. In our experiments, the reaction was monitored inside a demountable liquid cell (Harrick Scientific) consisting of two CaF$_{2}$ windows contained between the stainless steel housing and the cover plate. The length of the transmission path is set to 6 $\mu$m by inserting a PTFE spacer between the two windows.

\begin{figure}[h]
    \centering
    \includegraphics[scale=0.66]{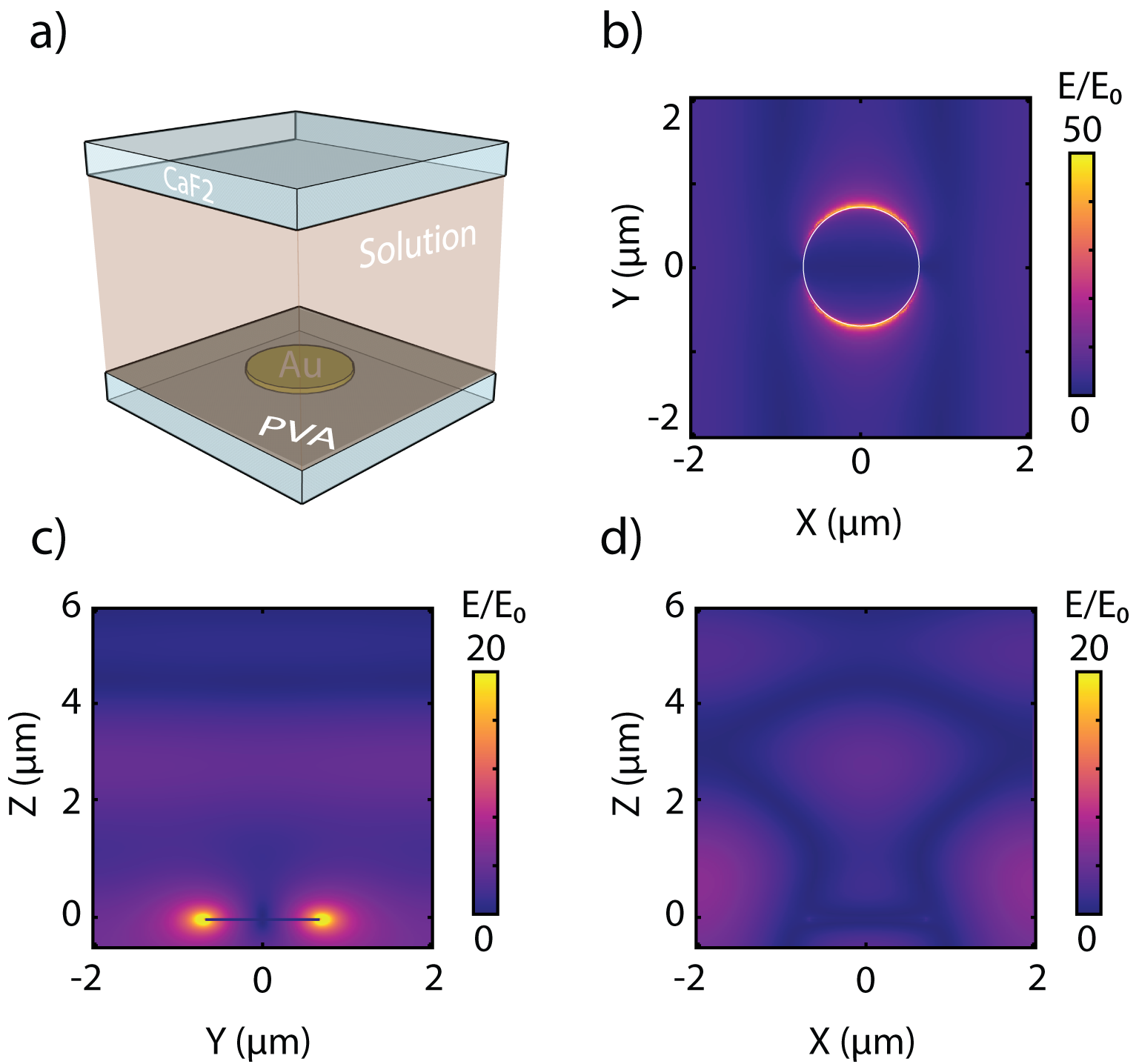}
    \caption{Simulations of the gold metasurface unit cell with the FDTD method. a) 3D schematic of the simulated unit cell. Electric-field enhancement cross-section at 1742 cm$^{-1}$ in the b) XY plane on top of the gold disk at 50 nm from the CaF$_{2}$, c) YZ plane, and d) XZ plane.}
    \label{fig:fig2}
\end{figure}

To facilitate the design and optimization of the non-local metasurface, we employ the finite-difference-in-time-domain method for characterizing the electric fields of SLRs (see details in SI section S2). The optimized geometric parameters include Au disks with 1.4 $\mu$m diameter, 50 nm height, in a square array with 4.05 $\mu$m period. A 3D representation of the simulated system is shown in Figure~\ref{fig:fig2}a. In Figure~\ref{fig:fig2}b - \ref{fig:fig2}d, we show the field enhancement ($\mathrm{E}/\mathrm{E}_0$) associated with the SLR on the unit cell of the metasurface. While the gold particle surface exhibits the highest field enhancement (Figure~\ref{fig:fig2}b and~\ref{fig:fig2}c), a significant portion of the SLR field extends into the liquid cell (Figure~\ref{fig:fig2}d). This property ensures effective coupling between a substantial number of molecules in the liquid reaction mixture and the SLR of the metasurface. 

\begin{figure}[h!]
    \centering
    \includegraphics[width=0.47\textwidth]{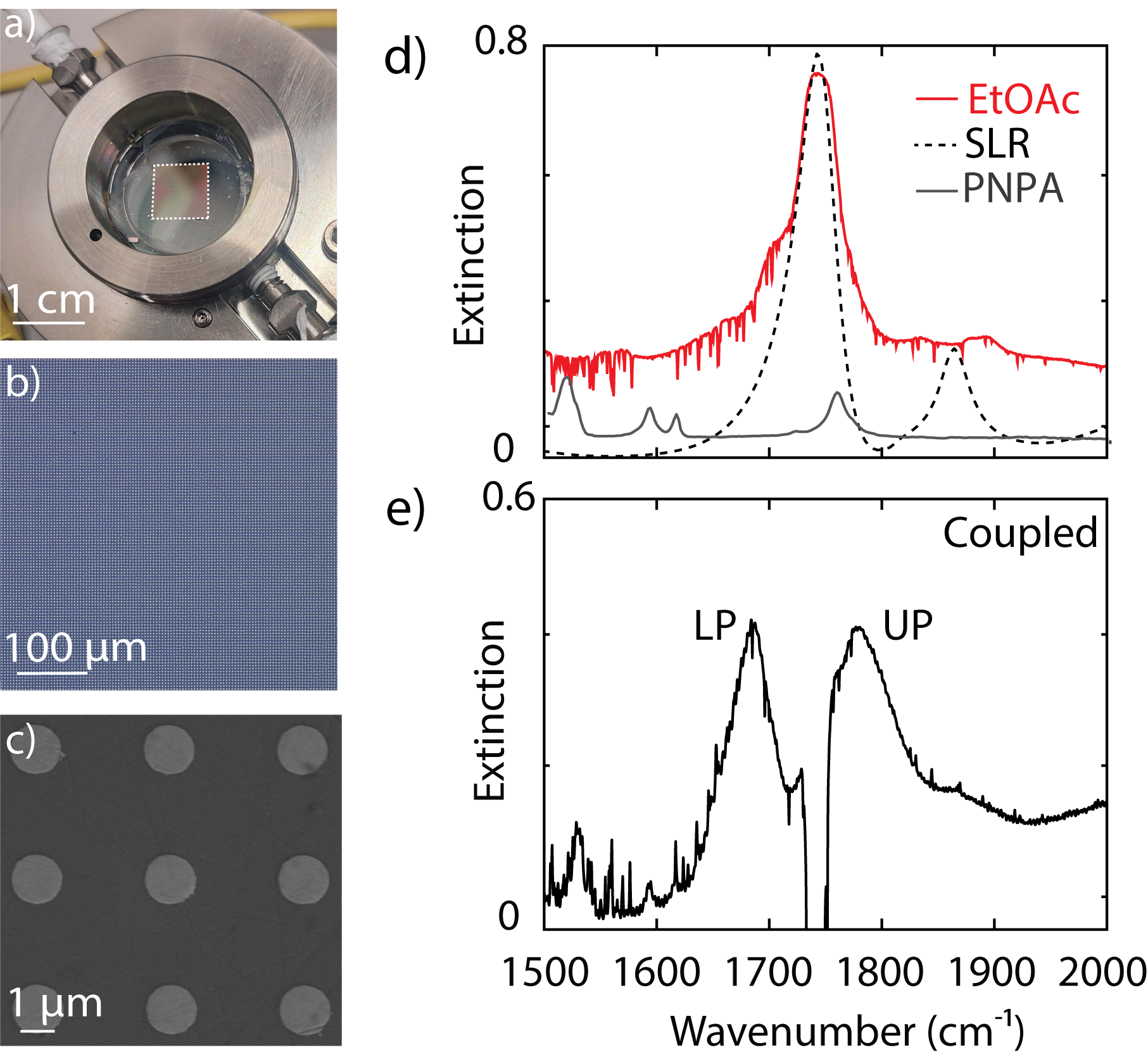}
    \caption{Metasurface cavity. a) Image of the liquid cell with the metasurface as one of the windows (the size of the array is indicated by the square). b) Optical image of the metasurface. c) Scanning electron microscope image of the gold disks forming the metasurface. d) Optical extinction spectra of the bare cavity showing the SLRs (black dotted), PNPA (gray solid) and EtOAc (red solid). e) Relative extinction spectrum of the coupled system. The y-axis of e) is normalized by one of the bare PNPA solutions to highlight the formation of polaritonic peaks (See raw data in SI section S5).
}
    \label{fig:fig3}
\end{figure}

The metasurface resonant with the molecular vibration is manufactured using electron beam lithography (EBL) on a CaF$_{2}$ substrate (see SI section S3). Furthermore, a detuned metasurface with a period of 4.3 $\mu$m and disk diameter of 1.01 $\mu$ m is fabricated, allowing us to examine the reaction in the presence of the metasurface without VSC.  Subsequently, we integrate the metasurface into the custom-designed liquid cell. As illustrated in Figure~\ref{fig:fig3}a, the array is located in the center of the liquid cell. The array is manufactured in an area of 9 mm $\times$ 9 mm. Optical microscope and scanning electron microscope images of the metasurface and the plasmonic microparticles are shown in Figure~\ref{fig:fig3}b and~\ref{fig:fig3}c, respectively.

With the liquid cell shown in Figure~\ref{fig:fig3}a, we characterize the C$=$O stretching mode of ethyl acetate without the metasurface through the extinction spectra (Figure~\ref{fig:fig3}d). By fitting the carbonyl peak with a Lorentzian lineshape (see SI Section S4), we extract the peak frequency 1742 cm$^{-1}$ and linewidth 50 cm$^{-1}$. In addition, we plot the simulated SLR spectrum of the metasurface (dotted black curve) to illustrate the spectral overlap with the measured molecular vibration.
The coupled system is characterized by measuring the extinction spectrum of the solution inside the liquid cell with the metasurface on one of the windows (Figure~\ref{fig:fig3}e). The result clearly shows the Rabi splitting, which is characteristic of VSC. We retrieve polariton energies of 1695 cm$^{-1}$ and 1772 cm$^{-1}$ for the lower and upper polaritons, respectively, with a Rabi splitting of 77 cm$^{-1}$ comparable to previous studies on VSC with polymer films.\cite{verdelli2022chasing}.

\begin{figure}[h!]
    \centering
    \includegraphics[width=0.47\textwidth]{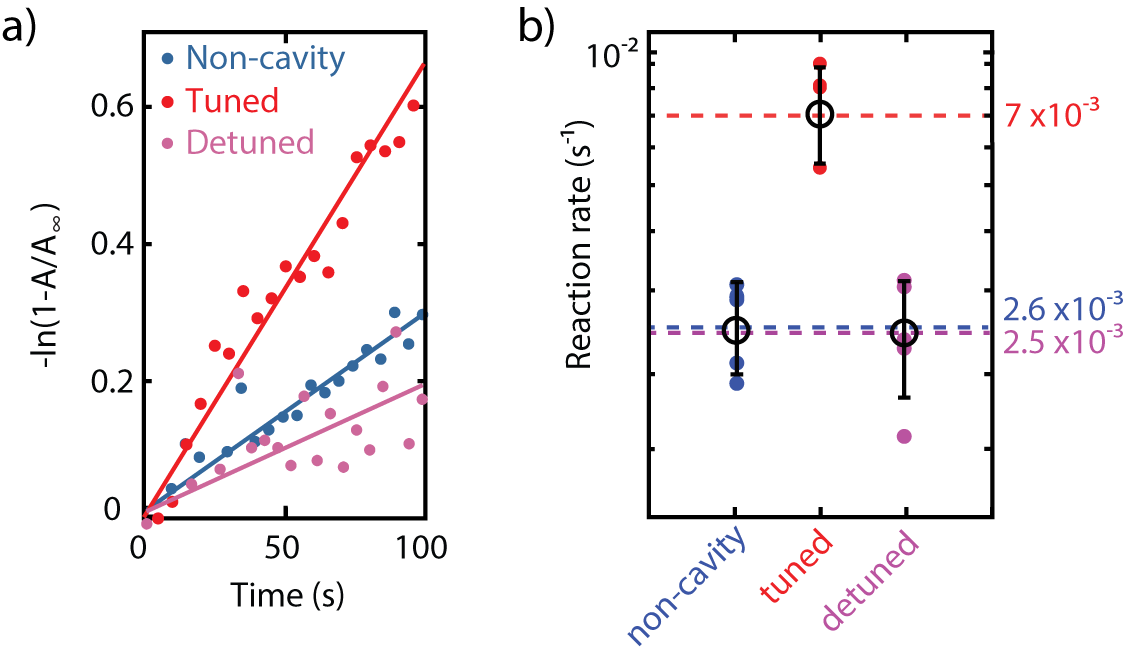}
    \caption{a) Kinetic traces of the reaction over time, showing measurements (filled circles) and linear fits (lines) for the non-cavity reaction (blue), the tuned metasurface reaction (red), and the detuned metasurface reaction (violet). b) Measured reaction rates for the non-cavity reaction (blue), the tuned metasurface reaction (red), and the detuned metasurface reaction (violet). The scattered solid circles are individual measurements, while the black open circles represent the average reaction rates, with error bars indicating the standard deviation of the different measurements.}
    \label{fig:fig4}
\end{figure}

Next, we investigate how the VSC caused by the SLR of the metasurface modifies the solvolysis reaction kinetics. A series of measurements are performed for the bare reaction without metasurface, the reaction with the metasurface tuned to the C$=$O of PNPA/EtOAc, and the reaction with the detuned metasurface (See details in SI sections S6-S9). In Figure~\ref{fig:fig4}a, the kinetics of the reaction in the different cases are investigated by analyzing the pseudo-first-order kinetic trace of the reaction expressed as $-\ln\left(1-\frac{A(t)}{A_{\infty}}\right)$, where $A_{\infty}$ is the absorbance at infinite time \cite{lather2019cavity}. The fitted rate constants in different cases are shown in Figure~\ref{fig:fig4}b. For the non-cavity reaction, we retrieve a reaction rate of $(2.6\pm0.4)\times$ 10$^{-3}$ s$^{-1}$, consistent with prior studies on para-nitrophenyl solvolysis.\cite{lather2019cavity}  Under VSC, the reaction rate significantly accelerates compared to the uncatalyzed reaction, reaching $(7.0\pm2.0)\times10^{-3}$ s$^{-1}$ during the initial 100 seconds, which is 2.7 times faster than the non-cavity reaction. In contrast, the control measurement involving a detuned metasurface results in a reaction rate of $(2.5\pm0.6)\times$ 10$^{-3}$ s$^{-1}$, demonstrating no enhancement compared to the non-cavity reaction. A clear comparison between these cases validates the pronounced acceleration of the reaction rates in the tuned metasurface. We speculate that VSC selectively weakens the carbonyl bond, rendering it more susceptible to nucleophilic attack by fluoride ions and thus increasing the substitution reaction rate.

For additional insight and comparison, we also conducted measurements of reaction kinetics in a tuned Fabry-Perot cavity with partially transmitting mirrors in the visible range, resulting in an increased reaction rate of $(3.9\pm1.3)\times$ 10$^{-3}$ s$^{-1}$ (See SI section S10), which aligns with a previous study.\cite{lather2019cavity} The reproducible rate enhancements observed in both the Fabry-Perot cavity and the plasmonic array validate the reliability of our experimental procedures. Notably, the different rate enhancements in Fabry-Perot and metasurface cavities may be attributed to their distinct spatial distributions of field enhancements. As illustrated in Figure~\ref{fig:fig2}, the metasurface exhibits a more localized field enhancement close to the array's surface, while the Fabry-Perot cavity field is distributed between the two mirrors (see SI section S10). These varying field distributions result in different catalytic effects on chemical reactions. Therefore, reducing the thickness of the cell for the metasurface should lead to a further increase in the measured reaction rate, as a larger fraction of probed molecules will be within the cavity field defined by the array. It is essential to note that reducing the cavity length of the Fabry-Perot cavity is not feasible, as this modification would alter its resonance frequency and detune it from the molecular vibration.

In summary, we have demonstrated the versatile application of metasurfaces as cavities in polaritonic chemistry, demonstrating for the first time that VSC induced by non-local resonances can significantly alter chemical reactions. Here, VSC is achieved by coupling SLRs supported by metasurfaces of Au microparticles and the C$=$O stretching mode of PNPA and EtOAc. Under VSC, we observed a 2.7-fold increase in the rate of the ester solvolysis reaction of PNPA compared to the non-cavity reaction. These findings pave the way for designing optical cavities that enable polaritonic chemical reactions through metasurfaces, which are particularly suitable for large-scale applications in flow reactors.

\begin{suppinfo}

Ester solvolysis reaction mechanism; design and optimization of the metasurface; sample fabrication; ethyl acetate Lorentzian fit;  characterization of the strongly coupled system; experimental methods of measuring reaction kinetics; spectroscopy of the non-cavity reaction, the reaction with the tuned/detuned metasurface and the reaction with the tuned Fabry-Perot cavity; comparison of field enhancements between the Fabry-Perot and the metasurface; comparison between the reactions rates.

\end{suppinfo}


\begin{acknowledgement}
We thank T.W. Ebbesen for useful discussions. J.G.R. acknowledges financial support from the Dutch Research Council (NWO) through the talent scheme (Vici Grant No. 680-47-628). A.B. acknowledges support from the Dutch Research Council (NWO) through the talent scheme (Vidi Grant No. 680-47-550). Y.-C.W. acknowledges support from the National Science and Technology Council (NSTC) through the postdoctoral research abroad program.
\end{acknowledgement}

\bibliography{reference}

\end{document}